\documentclass{pasj00}
\draft

\begin{document}
\SetRunningHead{Paron, S., Ortega, M. E., Astort, A., Rubio, M., Fari\~{n}a, C.}{The molecular ambient towards EGO G35.04-0.47}

\title{Studying the Molecular Ambient towards the Young Stellar Object EGO G35.04-0.47}

\author{Sergio \textsc{Paron},\altaffilmark{1,2}
Mart\'{\i}n \textsc{Ortega},\altaffilmark{1}
Ana \textsc{Astort},\altaffilmark{1,3}
M\'onica \textsc{Rubio},\altaffilmark{4}
\and
Cecilia \textsc{Fari\~{n}a},\altaffilmark{5}
}

\altaffiltext{1}{Instituto de Astronom\'{\i}a y F\'{\i}sica del Espacio (IAFE), CC 67, Suc. 28, 1428 Buenos Aires, Argentina}
\altaffiltext{2}{FADU and CBC, Universidad de Buenos Aires, Ciudad Universitaria, Buenos Aires, Argentina}
\altaffiltext{3}{Departamento de F\'{\i}sica, FCEN, Universidad de Buenos Aires, Ciudad Universitaria, Buenos Aires, Argentina}
\altaffiltext{4}{Departamento de Astronom\'{\i}a, Universidad de Chile, Casilla 36-D, Santiago, Chile}
\altaffiltext{5}{Isaac Newton Group of Telescopes, E-38700, La Palma, Spain}

\email{sparon@iafe.uba.ar}


\maketitle

\begin{abstract}
We are performing a systematic study of the interstellar medium around extended green objects (EGOs), likely
massive young stellar objects driving outflows. EGO G35.04-0.47 is located towards a dark cloud at the northern-west 
edge of an HII region. Recently, H$_{2}$ jets  were discovered towards this source, mainly towards its southwest, where the 
H$_{2}$ 1--0 S(1) emission peaks. Therefore, the source was catalogued as 
the Molecular Hydrogen emission-line object MHO 2429. 
In order to study the molecular ambient towards this star-forming site, we observed a region around the aforementioned EGO using the Atacama
Submillimeter Telescope Experiment in the $^{12}$CO J=3--2, $^{13}$CO J=3--2,
HCO$^{+}$ J=4--3, and CS J=7--6 lines with an angular and spectral resolution of 22\arcsec~and 0.11 km s$^{-1}$, respectively. 
The observations revealed a molecular clump where the EGO is embedded
at v$_{\rm LSR} \sim 51$ km s$^{-1}$, in coincidence with the velocity of a Class I 95 GHz methanol maser previously detected.
Analyzing the $^{12}$CO line we discovered high velocity molecular gas in the range from 34 to 47 km s$^{-1}$, most likely a blueshifted 
outflow driven by the EGO.
The alignment and shape of this molecular structure coincide with those of the southwest lobe of MHO 2429
mainly between 46 and 47 km s$^{-1}$, confirming that we are mapping its CO counterpart.
Performing a SED analysis of EGO G35.04-0.47 we found that its central object should be an intermediate-mass
young stellar object accreting mass at a rate similar to those found in some massive YSOs. We suggest that this source 
can become a massive YSO.
\end{abstract}

\section{Introduction}

More than 400 Galactic sources with extended emission at 4.5 $\mu$m were identified \citep{cyga08,chen13}. These objects were called 
extended green objects (EGOs) for the common coding of the 4.5 $\mu$m band as green in three-color
composite Infrared Array Camera images from the {\it Spitzer} Telescope.
According to the authors, an EGO is a probable massive young stellar object (MYSO) driving outflows.
The extended emission in the 4.5 $\mu$m band is supposed to be due to H$_{2}$ ($\nu=0-0$, S(9,10,11))
lines and CO ($\nu = 1-0$) band heads,
that are excited by the shock of the outflows propagating in the interstellar medium (ISM) (e.g. \cite{noriega04}). 
Some studies have shown that EGOs indeed trace outflows related to MYSOs \citep{debuizer10,cyga11}. On the other hand, 
based on the lack of detection of the shock-excited H$_{2}$ line at 2.122 $\mu$m towards several EGOs, \citet{lee12} suggest 
that the 4.5 $\mu$m emission can also be due to scattered light.
Studying the molecular ambient towards EGOs is useful not only to probe the relationship between the 4.5 $\mu$m extended
emission and shocked gas but also to study how YSOs can disrupt their surrounding environment,
through the clearing of gas and the injection of momentum and energy onto the medium.
Indeed, bipolar jets from protostars appear to be ubiquitous and to have a close relationship to the accretion
process from a disc into a forming star  \citep{froebrich03}.
These jets can excavate cavities by sweeping up
the molecular material which then forms an outflow that can be detected through the emission of CO lines.

We are performing a systematic study of the ISM around EGOs in order to find evidence of outflowing
activity and to characterize the molecular ambient where the source is embedded. In this paper 
we present the results from the study of EGO G35.04-047 

\begin{figure}
\begin{center}
\FigureFile(150mm,150mm){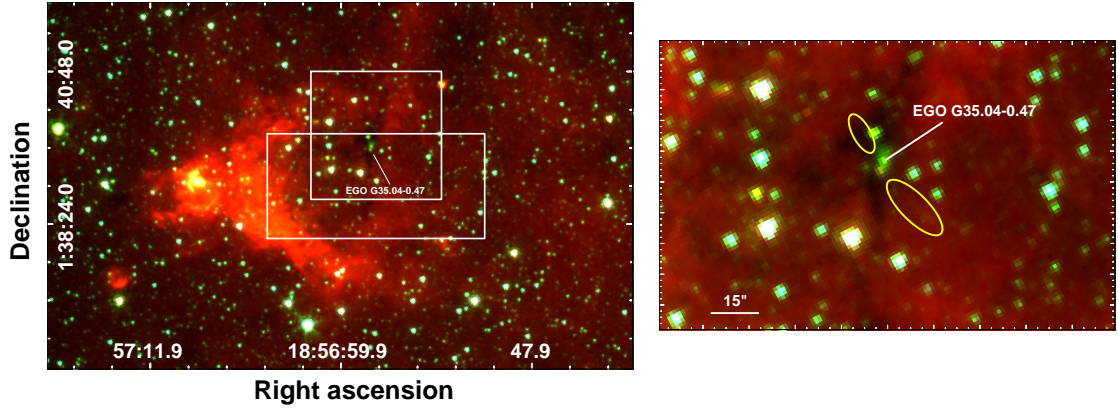}
\end{center}
\caption{Left: Three-colour {\it Spitzer}-IRAC image of the region where EGO G35.04-0.47 is located
(8 $\mu$m = red, 4.5 $\mu$m = green, and 3.5 $\mu$m = blue).
The $^{12}$CO J=3--2 and HCO$^{+}$ J=4--3 observations covers the white rectangle, while the observations of $^{13}$CO J=3--2 and CS J=7--6, 
the square.
Right: a zoom towards the EGO position.
The yellow ellipses represent schematically the lobes of MHO 2429 according to the H$_{2}$ 1--0 S(1)
image presented in \citet{lee12}.}
\label{present}
\end{figure}

The source EGO G35.04-047 lays in a border of the HII region G035.0-0.5. 
Figure \ref{present}  is a three-color {\it Spitzer}-IRAC composite image of the field, where the HII region G035.0-0.5 appears 
conspicuous at the 8 $\mu$m
emission and the EGO G35.04-0.47, located in a dark cavity at the northern-west edge of the HII region, can be distinguished by its 
4.5 $\mu$m green emission. The  
HII region has a systemic velocity  of v$_{\rm LSR} \sim$ 51 km s$^{-1}$~\citep{lockman89,kuchar97}.
EGO G35.04-0.47 is very likely embedded in the dark cloud 
SDC G35.041-0.471 \citep{peretto09} suggesting that it could be an YSO evolving in its parental dark and dense cloud.
Moreover, it is probable that the formation of this YSO has been triggered by 
the expansion of the HII region G035.0-0.5. It is likely that the shock front of the HII region has overrun and compressed 
a preexisting molecular clump leading the formation of the YSO (e.g. \cite{paron11,diri12}).
\citet{chen11} detected Class I 95 GHz methanol maser at the center of this EGO with a main component centered at 51.5 km s$^{-1}$, 
in agreement with a previous single-pointing study using the HCO$^{+}$, $^{12}$CO, $^{13}$CO, and C$^{18}$O J=1--0 lines \citep{chen10}. 
More recently, \citet{he12} also reported a v$_{\rm LSR} \sim$ 51 km s$^{-1}$ for EGO G35.04-0.47 from mm molecular line observations.
\citet{cyga2013} detected NH$_{3}$ towards this EGO at
v$_{\rm LSR} \sim 51.15$ km s$^{-1}$, reporting a distance of 3.11$\pm$0.40 kpc in agreement with the distance of 3.4 kpc 
reported in \citet{chen11}.
In particular, this EGO shows near-IR emission extending mainly towards the southwest. 
\citet{lee12} using near-IR data from the UKIDSS \citep{law07} confirmed the presence of a bipolar H$_{2}$ jet detected at 2.122 $\mu$m, with a strong 
component towards the southwest of the EGO and not aligned with the 4.5 $\mu$m emission. The yellow ellipses in figure \ref{present}
(right) schematically represent the lobes of the bipolar H$_{2}$ jet. 
The authors catalogued this object as the Molecular Hydrogen emission-line 
object MHO 2429\footnote{MHO catalogue is hosted by Liverpool John Moores University}. MHOs are spatially resolved objects associated with 
outflows from young stars \citep{davis10}.

Based on new submillimeter molecular observations acquired with the Atacama Submillimeter Telescope Experiment (ASTE), 
we present a study of the ISM towards EGO G35.04-0.47 in order to characterize the molecular gas
probably related to the star-forming processes. We mainly analyze the ISM southwards the EGO with the goal 
of confirming the existence of a molecular outflow related to the strong southwestern H$_{2}$ jet.
Additionally, we study the nature of the driving source.

\section{Observations and Data Reduction}
\label{obs}

The molecular observations presented in this work were carried out on June 12 and 13, 2011
with the 10 m Atacama Submillimeter
Telescope Experiment (ASTE; \cite{ezawa04}). We used the CATS345 GHz band receiver, which is a two-single
band SIS receiver remotely tunable in the LO frequency range of 324-372 GHz. 
We observed $^{13}$CO J=3--2 at 330.588 GHz and CS J=7--6 at 342.883 GHz mapping a region
of 2\farcm0~$\times$ 2\farcm0~centered at
RA $=$\timeform{18h56m57.9s}, dec. $=+$\timeform{01D39'47.7''}, J2000 (white square in left panel of figure \ref{present}).
Additionally, we simultaneously observed $^{12}$CO J=3--2 at 345.796 GHz and HCO$^{+}$~J=4--3 at
356.734 GHz, mapping a region of 3\farcm4~$\times$ 1\farcm6 centered at RA $=$ \timeform{18h56m57.8s},
dec. $=+$\timeform{01D38'59.8''}, J2000 (white rectangle in left panel of figure \ref{present}). 
The mapping grid spacing was 20\arcsec~in
all cases and the integration time was 60 sec in each pointing for the first set of data and 30 sec for the second one. 
All the observations were performed in position switching mode
and the half-power beamwidth (HPBW) was about 22\arcsec.

We used the XF digital spectrometer with bandwidth and spectral resolution set to
128 MHz and 125 kHz, respectively, corresponding to a velocity resolution of about 0.11 km s$^{-1}$.
The weather conditions were optimal and the system temperature
varied from T$_{\rm sys} = 150$ to 200 K. 
The absolute pointing accuracy and main beam efficiency were verified by observing the
$^{12}$CO J=3--2 emission towards the standard source V1427Aq (RA $=$\timeform{19h13m58.6s},dec. $=+$\timeform{00D07'31.9''}, J2000, with
an antenna temperature of T$_{\rm a} = 0.2$ K) and
were about 2\arcsec~and $\eta_{\rm mb} \sim 0.65$, respectively.
The spectra were Hanning smoothed to improve the signal-to-noise ratio and only linear or some second order
polynomia were used for baseline fitting. The rms noise of the spectra is between 0.15 and 0.20 K.
The data were reduced with NEWSTAR\footnote{Reduction software based on AIPS developed at NRAO,
extended to treat single dish data
with a graphical user interface (GUI).} and the spectra processed using the XSpec software
package\footnote{XSpec is a spectral line reduction package for astronomy which has been
developed by Per Bergman at Onsala Space Observatory.}.

\section{Results and Discussion}
\label{results}

\begin{figure}
\begin{center}
\FigureFile(64mm,65mm){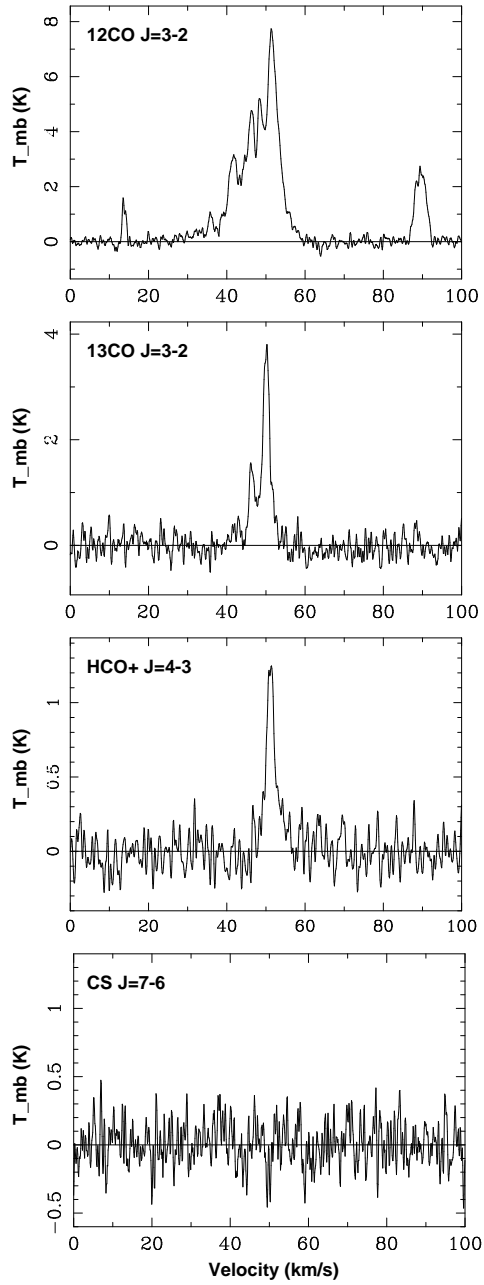}
\end{center}
\caption{$^{12}$CO and $^{13}$CO J=3--2, HCO$^+$ J=4--3, and CS J=7--6 spectra observed towards the EGO G35.04-0.47 position. }
\label{spectra}
\end{figure}

\begin{figure}
\begin{center}
\FigureFile(100mm,100mm){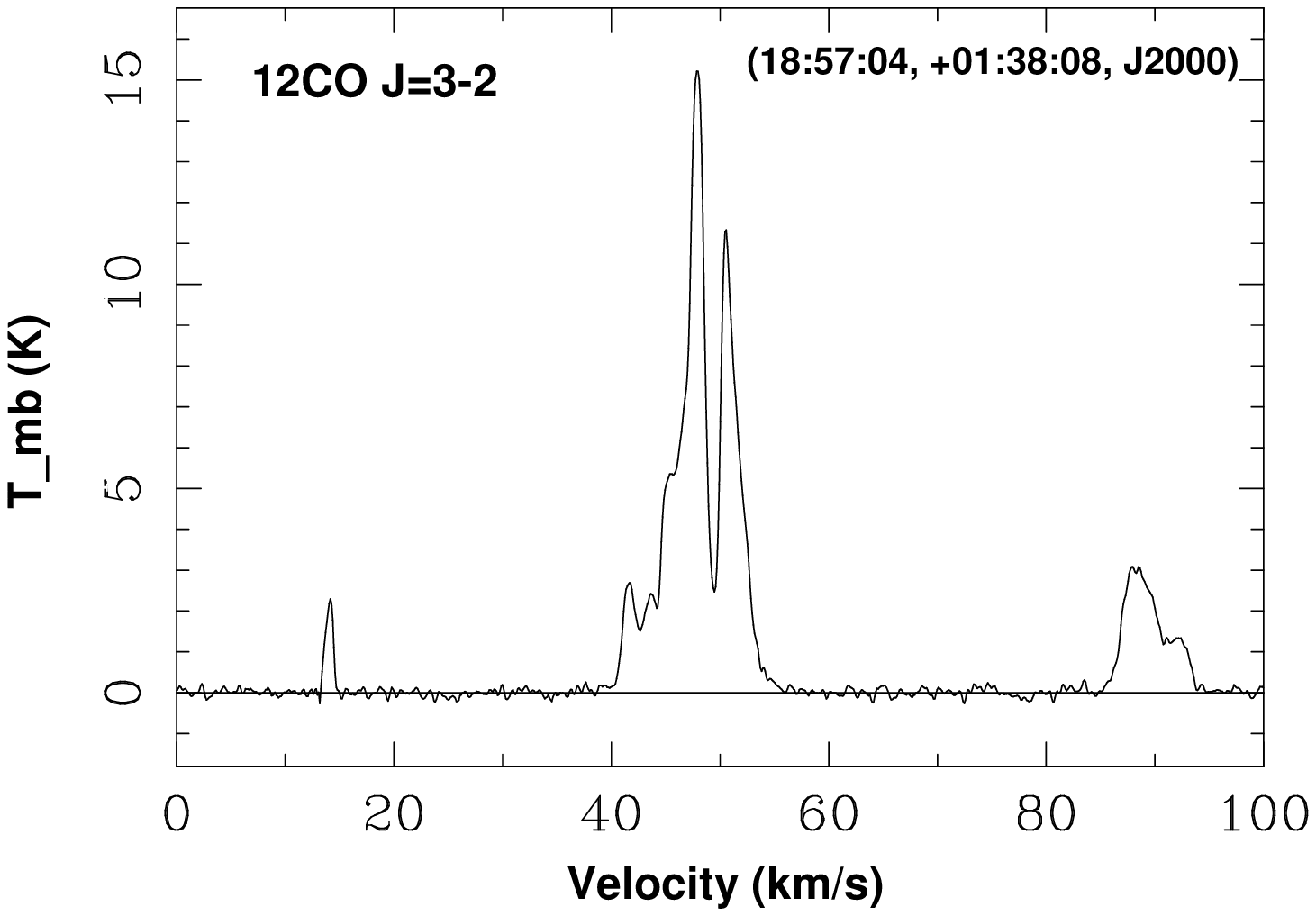}
\end{center}
\caption{$^{12}$CO J=3--2 spectrum obtained at RA $=$\timeform{18h57m04s}, dec. $=+$\timeform{01D38'08''}, J2000,
a position $\sim$ 2.5 arcmin far from the EGO.   }
\label{spectrumFar}
\end{figure}

Figure \ref{spectra} shows the spectra of the four different molecular species ($^{12}$CO J=3--2, $^{13}$CO J=3--2, HCO$^{+}$ J=4--3, and CS J=7--6) 
observed towards the EGO G35.04-0.47 position.
Emission from the CS J=7--6 line was not detected in the whole surveyed field. The $^{12}$CO J=3--2 spectrum shows a complex 
behaviour with a main component centered at v $\sim 51$ km s$^{-1}$~and very likely related components towards lower velocities 
(between 34 and 50 km s$^{-1}$).
This spectrum also presents less intense components centered at $\sim$ 14 and
89 km s$^{-1}$. Taking into account that these molecular components appear in the whole observed area
we discard the possibility that they represent CO bullets related to the YSO seen along the line 
of sight. To illustrate this, figure \ref{spectrumFar} displays the $^{12}$CO J=3--2 spectrum at 
RA $=$\timeform{18h57m04s}, dec. $=+$\timeform{01D38'08''}, J2000, a position $\sim$2.5 arcmin far from the EGO (at the bottom left
corner of the $^{12}$CO field shown in the left panel of figure \ref{present}), where the components at 
$\sim$ 14 and 89 km s$^{-1}$~are also present. Thus, we conclude that 
they are unrelated molecular components seen along the line of sight. Therefore, our analysis focus on the components 
that appear between 34 and 60 km s$^{-1}$. Within this velocity range, as shown in figure \ref{spectra}, the $^{12}$CO spectrum obtained at 
the EGO position has a very likely wing towards lower velocities with several components. This may be due to shocks 
and/or turbulent motion in the gas (see e.g. \cite{falgarone94}). It is probable that some of these components arise from a blueshifted outflow
related to the southwestern H$_{2}$ structure found by \citet{lee12}. 
This issue will be further studied in section \ref{mapping}.
While the $^{12}$CO J=3--2 profile exhibits a complex structure, the $^{13}$CO and HCO$^{+}$ spectra, on the other hand, show a simpler behaviour.
In the case of the $^{13}$CO J=3--2 emission, the profile has two components centered at $\sim$ 50.1 and 46.5 km s$^{-1}$, while the HCO$^{+}$ i
spectrum has only one component, centered at $\sim 51.2$ km s$^{-1}$. In table \ref{lines} we present the parameters determined from the profile Gaussian fitting, 
where T$_{mb}$ represents the peak brightness temperature, V$_{\rm LSR}$ the central velocity referred to the Local Standard of Rest, 
and $\Delta$v the FWHM line width. Errors are formal 1$\sigma$ value for the model of the Gaussian line shape. 
The parameters derived for the  HCO$^{+}$ J=4--3 line are very similar to those obtained by \citet{schlin11} for the HCO$^{+}$ J=3--2 towards 
this region. 

\begin{figure}
\begin{center}
\FigureFile(95mm,95mm){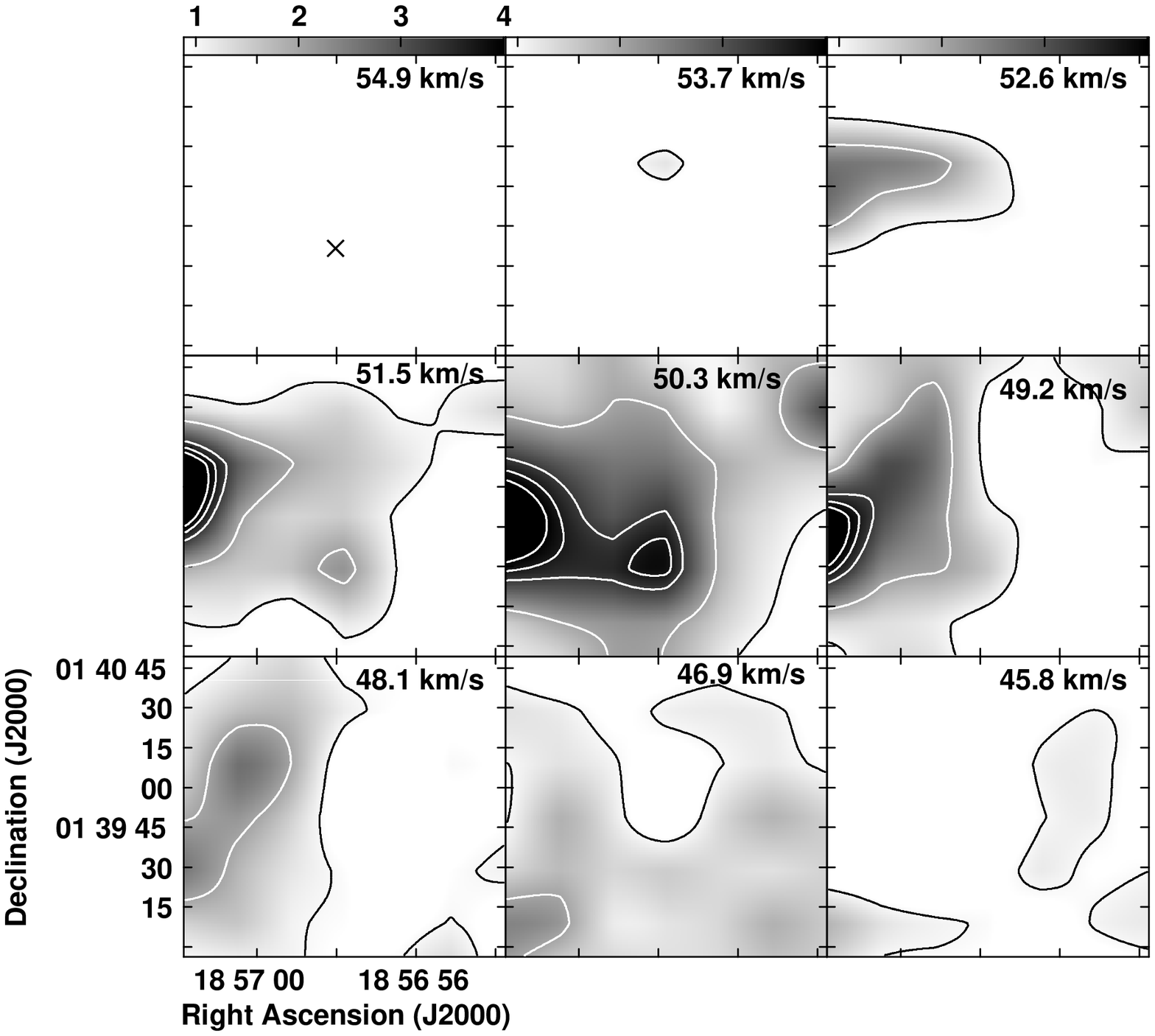}
\end{center}
\caption{Channel maps of the $^{13}$CO J=3--2 emission integrated in steps of about 1.1 km s$^{-1}$.
The grayscale is in K km s$^{-1}$~and the contour levels are 1.0, 2.0, 3.3, 3.7, and 4.0 K km s$^{-1}$.
The cross in the first panel indicates the EGO G35.04-0.47 position.}
\label{13copanels}
\end{figure}

The morphology and velocity distribution of the molecular gas around EGO G35.04-0.47 can be seen 
in figure \ref{13copanels}, where the $^{13}$CO J=3--2 emission is presented in a series of channel maps integrated 
in steps of about 1.1 km s$^{-1}$.
Most of the panels show molecular gas related to the HII region G035.0-0.5, mainly towards the east. The panel
at 50.3 km s$^{-1}$~shows a molecular clump, almost at the field center, whose maximum coincides with the position of the EGO. 
In the panel at 46.9 km s$^{-1}$~it can be seen a cavity northward the EGO position. 

\begin{figure}
\begin{center}
\FigureFile(130mm,130mm){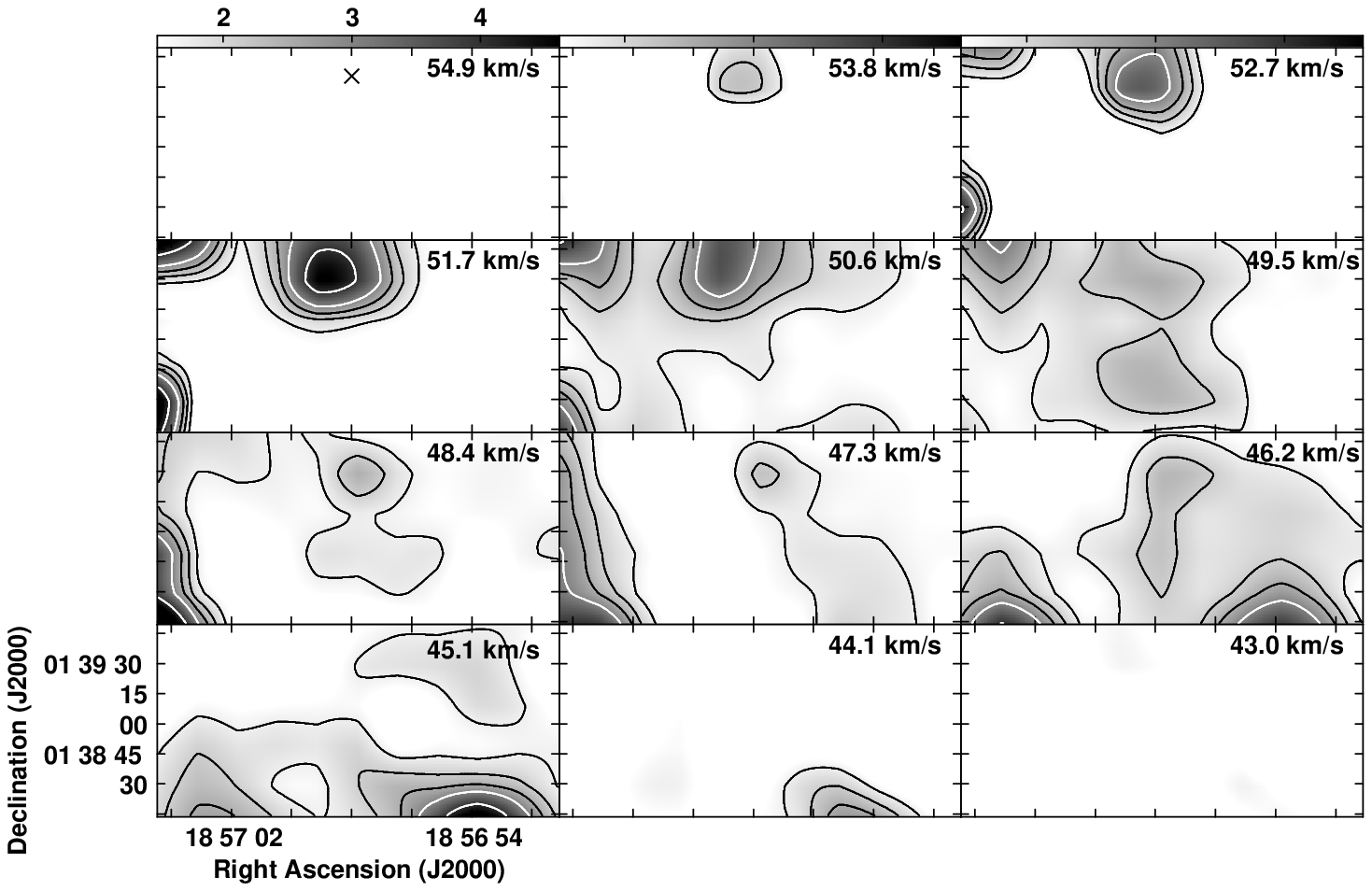}
\end{center}
\caption{Channel maps of the $^{12}$CO J=3--2 emission integrated in steps of about 1.1 km s$^{-1}$.
The grayscale is in K km s$^{-1}$~and the contour levels are 1.7, 2.1, 2.5, 3.0, and 4.0 K km s$^{-1}$.
The cross in the first panel indicates the EGO G35.04-0.47 position.}
\label{chann1}
\end{figure}

\begin{figure}
\begin{center}
\FigureFile(150mm,150mm){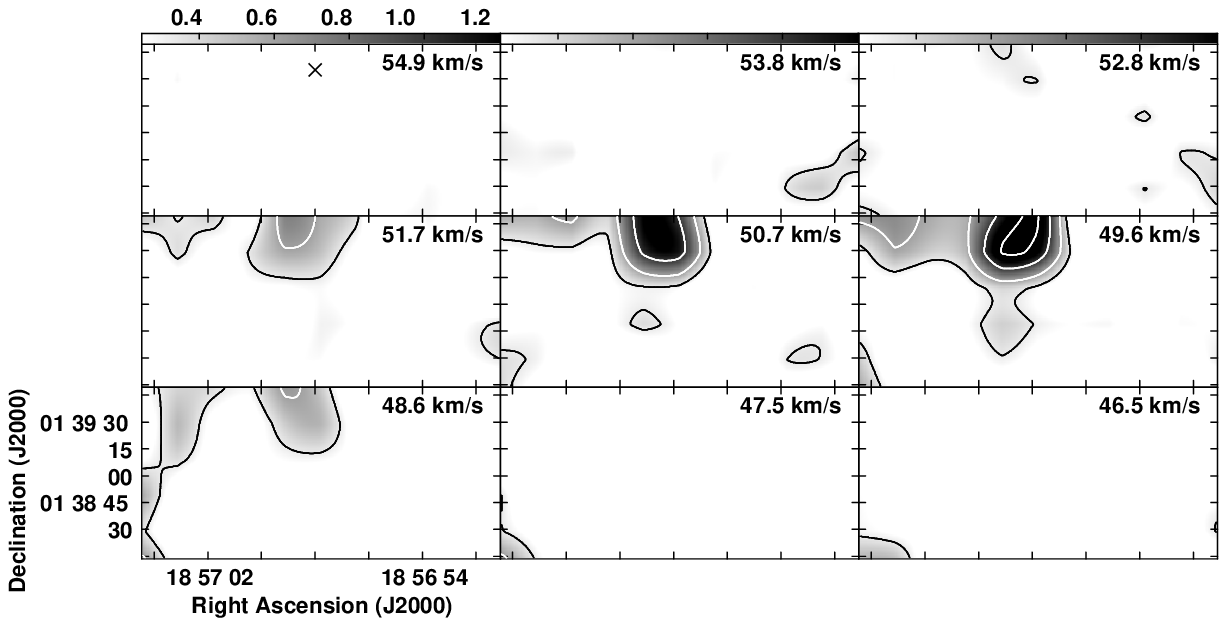}
\end{center}
\caption{Channel maps of the HCO$^{+}$ J=4--3 emission integrated in steps of about 1.1 km s$^{-1}$.
The grayscale is in K km s$^{-1}$~and the contour levels are 0.3, 0.6, 1.0, and 1.4 K km s$^{-1}$.
The cross in the first panel indicates the EGO G35.04-0.47 position.}
\label{hco+panels}
\end{figure}

The distribution of the $^{12}$CO J=3--2 emission, tracer of molecular outflows, towards the south of the EGO 
can be observed in figure \ref{chann1}. Note that the strong H$_{2}$ jet mapped by \citet{lee12} is located towards the southwestern region of the EGO 
(see right panel of figure \ref{present}).
The $^{12}$CO J=3--2 emission is presented in figure \ref{chann1} in a series of channel maps integrated in steps of about 1.1 km s$^{-1}$. The position of
EGO G35.04-0.47 coincides with the molecular clump that appears northwards in the panels from 53.8 to 50.6 km s$^{-1}$, peaking at
about 51 km s$^{-1}$. At lower velocities, see panels between  48.4 and 46.2 km s$^{-1}$, there is an elongated molecular feature 
aligned with the southwestern H$_{2}$ jet mapped by \citet{lee12} (see mainly
panel at 47.3 km s$^{-1}$). This molecular feature is connected towards the southwest with an unrelated molecular clump, not 
completely mapped, whose emission peaks between 45.1 and 46.2 km s$^{-1}$. The molecular gas that
appears eastwards is related to the HII region G035.0-0.5.
Figure \ref{hco+panels} exhibits a similar plot for the HCO$^{+}$ J=4--3 emission.
It is important to note that the central velocity of the molecular clump ($\sim51$ km s$^{-1}$) mapped in the three lines and related to the 
EGO is in agreement with that of the Class I 95 GHz methanol maser detected by \citet{chen11} and the NH$_{3}$ emission
detected by \citet{cyga2013}, suggesting to be the systemic velocity of the parental cloud where the EGO is embedded.

\subsection{Column Densities and Abundances}
\label{colabs}

In order to have a rough estimate of the molecular column densities and abundances of the clump where EGO G35.04-0.47 is embedded, 
we assume local thermodynamic equilibrium (LTE) and a beam filling factor of 1.
We calculate the excitation temperature from
\begin{equation}
T_{ex}(3 \rightarrow  2) = \frac{16.59 {\rm K}}{{\rm ln}[1 + 16.59 {\rm K} / (T_{\rm max}(^{12}{\rm CO}) + 0.036 {\rm K})]}
\label{eq2}
\end{equation}
where $T_{\rm max}(^{12}{\rm CO}$) is the $^{12}$CO peak temperature of the component centered at $\sim51$ km s$^{-1}$, 
obtaining $T_{ex} \sim 13.5$ K. 
Then, assuming that the $^{13}$CO is optically thin, we calculate its column density from:
\begin{equation}
{\rm N(^{13}CO)} = 8.28 \times 10^{13}~e^{\frac{15.87}{T_{ex}}}\frac{T_{ex} + 0.88}{1 - exp(\frac{-15.87}{T_{ex}})} \frac{1}{J(T_{ex}) - J(T_{\rm BG})} \int{{\rm T_{mb} ~dv}} 
\label{eq3}
\end{equation}
with
\begin{equation}
J(T) = \frac{h\nu/k}{exp(\frac{h\nu}{kT}) - 1},
\label{eq6}
\end{equation}
yielding N($^{13}$CO) $\sim 3 \times 10^{17}$ cm$^{-2}$. According to the Galactic position of this source and its distance (about 3.4 kpc), 
following \citet{milam05} we assume an isotope abundance ratio $X =$ [$^{12}$CO]/[$^{13}$CO] of 50 to obtain 
N($^{12}$CO) $\sim 1.5 \times 10^{19}$ cm$^{-2}$. 

The HCO$^{+}$ column density was derived from:
\begin{equation}
{\rm N(HCO^{+})} = 5.85 \times 10^{10}~e^{\frac{25.7}{T_{ex}}}\frac{T_{ex} + 0.71}{1 - exp(\frac{-17.12}{T_{ex}})}  \frac{1}{J(T_{ex}) - J(T_{\rm BG})} \frac{\tau}{1-e^{-\tau}} \int{{\rm T_{mb} ~dv}}
\label{eq8}
\end{equation}
and by considering that the HCO$^{+}$ J=4--3 is optically thick, we use a `cannonical' optical depth of $\tau = 1$  
and an excitation temperature of 43 K, which corresponds to the equivalent temperature of the molecular transition,
to obtain N(HCO$^{+}$) $\sim 3.2 \times 10^{12}$ cm$^{-2}$.

To independently estimate a value for the H$_{2}$ column density, we use the millimeter continuum flux at 1.1 mm measured in the Bolocam 
Galactic Plane Survey towards the source BGPS G35.045-0.478 \citep{roso10}, which is related to the molecular clump 
analyzed in this study. 
Using N(H$_{2}) = 2.19 \times 10^{22}[e^{13/T_{d}} - 1] S_{1.1}$ cm$^{-2}$ 
\citep{bally10} with the flux density obtained from the 40\arcsec~aperture (S$_{1.1} = 0.437$ Jy), which seems appropriate for the dimensions of the molecular clump, 
considering a dust temperature of $T_{d} =$ 13 K, and applying the multiplicative correction factors of 1.46 and 1.5 for the point source aperture and flux
density, respectively \citep{aguirre11}, we derive N(H$_{2}) \sim 3.5 \times 10^{22}$ cm$^{-2}$.
Then, using this value, we obtain the following molecular abundances:
X($^{12}$CO) $\sim 4.2 \times 10^{-4}$, X($^{13}$CO) $\sim 8.5 \times 10^{-6}$, and X(HCO$^{+}$) $\sim 1 \times 10^{-10}$.
The derived $^{12}$CO and $^{13}$CO abundances are in agreement with the `standard' value for dark clouds (e.g. \cite{vandish99}), while
HCO$^{+}$ abundance is similar to the values obtained towards high-mass star-forming regions \citep{cortes10,cortes11,ortega12}.

\subsection{Mapping a CO Outflow}
\label{mapping}

\begin{figure}
\begin{center}
\FigureFile(120mm,120mm){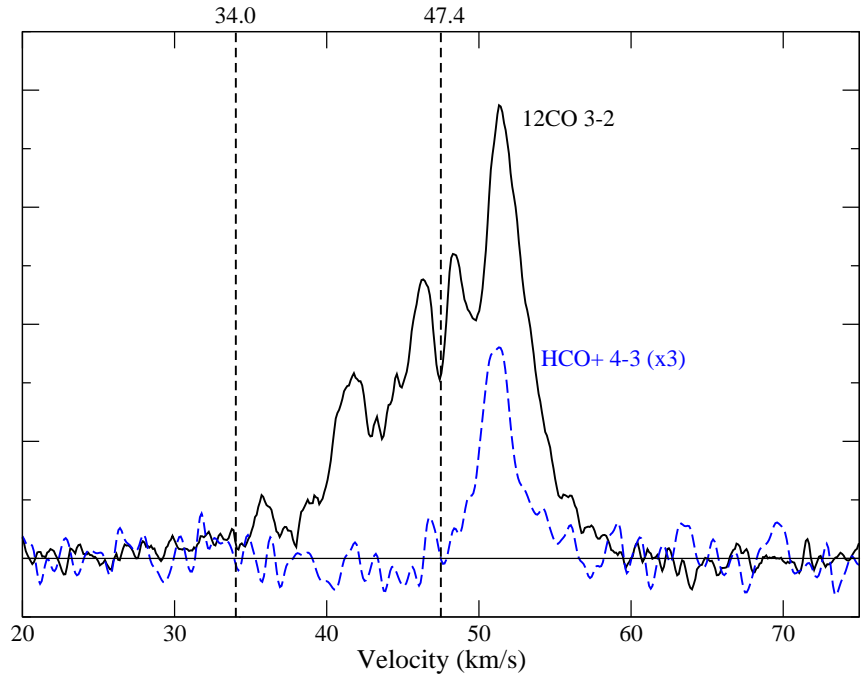}
\end{center}
\caption{$^{12}$CO J=3--2 (solid line) and HCO$^{+}$ J=4--3 scaled by a factor of $\times3$ (dashed line) profiles towards EGO G35.04-0.47.
The vertical lines indicate the range where $^{12}$CO emission is detected at lower velocities with respect to the HCO$^{+}$.}
\label{compara}
\end{figure}

Given that the $^{12}$CO is a good tracer of molecular outflows towards YSOs (e.g. \cite{beuther02,wu04}), we 
carefully analyze this line in order to find a  possible counterpart for the southwestern H$_{2}$ jet mapped by \citet{lee12}.
As previously mentioned, the $^{12}$CO spectrum displayed in figure \ref{spectra} shows 
a likely wing towards lower velocities with several components, being probable that some of them arise from a blueshifted outflow
related to the southwestern H$_{2}$ structure.
To confirm the presence of outflows or high velocity material moving along the line of sight we compare
the $^{12}$CO emission with the higher density tracer line HCO$^{+}$ J=4--3. Figure \ref{compara} presents the $^{12}$CO and HCO$^{+}$ spectra obtained 
towards the EGO G35.04-0.47 position. The vertical lines indicate the range where $^{12}$CO emission is detected at lower velocities with 
respect to the HCO$^{+}$, showing the presence of a $^{12}$CO spectral wing, which may be due to a blueshifted outflow extending from 34.0 to 47.4 km s$^{-1}$.
It is probable that within this velocity range there is also some molecular emission unrelated to the outflow activity and likely associated 
with the extended HII region and/or to the dark cloud. Thus, the derived outflow parameters using this range must be considered as upper 
limits. On the other side, none signature of gas moving at higher velocities (i.e. a probable redshifted outflow) is observed.

\begin{figure}
\begin{center}
\FigureFile(130mm,130mm){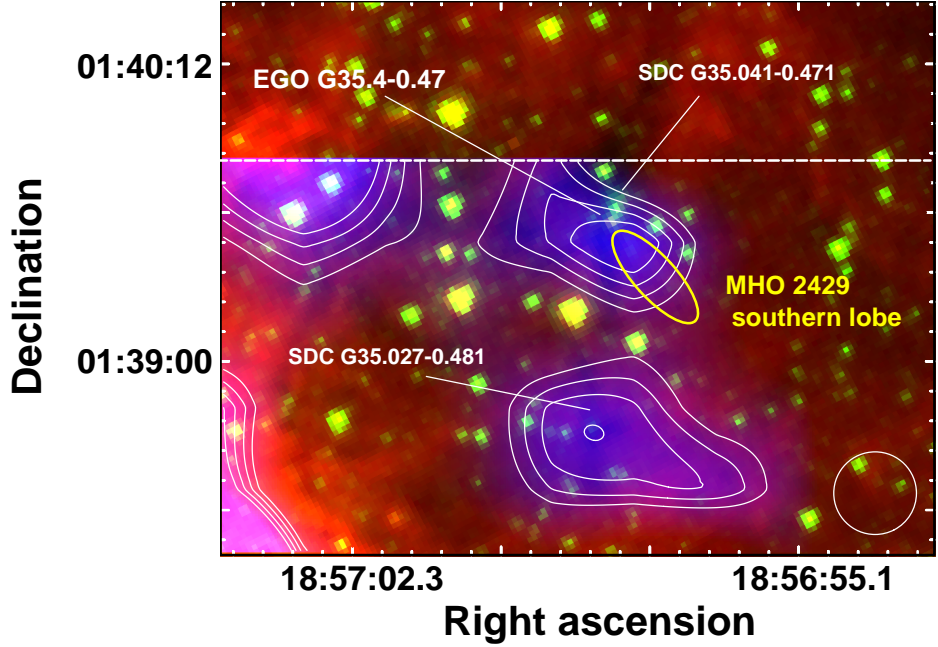}
\end{center}
\caption{Three colour image where the {\it Spitzer}-IRAC 4.5 and 8 $\mu$m are displayed in green and red, respectively,
and the $^{12}$CO J=3--2 emission integrated between 34.0 and 47.4 km s$^{-1}$~is shown in blue with white contours. The contour levels
are 6.9, 7.2, 7.5, and 7.8 K km s$^{-1}$.
The yellow ellipse represents schematically the southwestern lobe of MHO 2429 according to the H$_{2}$ 1--0 S(1) image presented
in \citet{lee12}. The Spitzer dark clouds catalogued in \citet{peretto09} are indicated. 
The beam of the $^{12}$CO observation is included in the bottom right corner and the dashed horizontal line is the upper
boundary of the $^{12}$CO observation.}
\label{outflowAll}
\end{figure}

Figure \ref{outflowAll} shows the $^{12}$CO J=3--2 emission integrated between 34.0 and 47.4 km s$^{-1}$~(in blue with white contours). Two molecular clumps 
peak in this velocity range, one towards the EGO G35.04-0.47 position, and the other
one, located towards the south, probably related to the dark cloud SDC G35.027-0.481 \citep{peretto09}. 
The $^{12}$CO emission towards the left side of the image is very likely related to the HII region border.
In order to roughly estimate the outflow mass, we again assume LTE and 
derive the $^{12}$CO optical depth $\tau_{12}$ from (e.g. \cite{scoville86}):
\begin{equation}
\frac{^{12}{\rm T}_{mb}}{^{13}{\rm T}_{mb}} = \frac{1-exp(-\tau_{12})}{1-exp(-\tau_{12}/X)},
\label{eq1}
\end{equation}
where $X = 50$ is the isotope abundance ratio.
We use the peak temperature ratio between the CO isotopes obtained from the components centered at $\sim46$ km s$^{-1}$, deriving
$\tau_{12} \sim 19$. In this case, equation (\ref{eq2}) gives an excitation temperature of T$_{\rm ex} \sim 10$ K.
Finally, integrating between 34.0 and 47.4 km s$^{-1}$ the following equation: 
\begin{equation}
{\rm N(^{12}CO)} = 7.96 \times 10^{13}~e^{\frac{16.6}{T_{ex}}}\frac{T_{ex} + 0.92}{1 - exp(\frac{-16.6}{T_{ex}})} \frac{1}{J(T_{ex}) - J(T_{\rm BG})} \frac{\tau}{1-e^{-\tau}} \int{{\rm T_{mb} ~dv}}
\label{eq4}
\end{equation}
we obtain a total column density N($^{12}$CO) $\sim 4 \times 10^{17}$ cm$^{-2}$.
This value results from the summation of column densities calculated at two different positions: two separated beams along the CO
structure related to the EGO. Then, we use
\begin{equation}
{\rm M} = \mu ~ {\rm m_{H_2}} ~ X_{CO}^{-1} ~ Area ~ d^{2} ~ {\rm N(^{12}{\rm CO})} 
\label{eqmass}
\end{equation}
to derive the total mass of the CO outflow. In this equation
$\mu$ is the mean molecular weight which was assumed to be 2.8 by taking into account a relative helium abundance
of 25 \%, m$_{\rm H}$ is the hydrogen mass, $X_{CO} = 10^{-4}$, $d$ is the distance assumed to be 3.4 kpc, and
the area of the molecular structure was approximated with an ellipse with semi-axes of 20\arcsec~and 12\arcsec. The total mass for the CO
outflow calculated in this way is 21 M$_\odot$.
Then we obtain a momentum of P $\sim 270~\times~{\rm cos}^{-1}(\phi)$ M$_\odot$~km s$^{-1}$, and an energy of
E $\sim 3.5 \times 10^{46}~ \times~{\rm cos}^{-2}(\phi)$ erg, where $\phi$ is the inclination angle of the outflow, which is uncertain.
As explained above, these values must be considered upper limits since it is possible that some $^{12}$CO emission towards the EGO is not
actually related with the outflow activity.

\begin{figure}
\begin{center}
\FigureFile(110mm,110mm){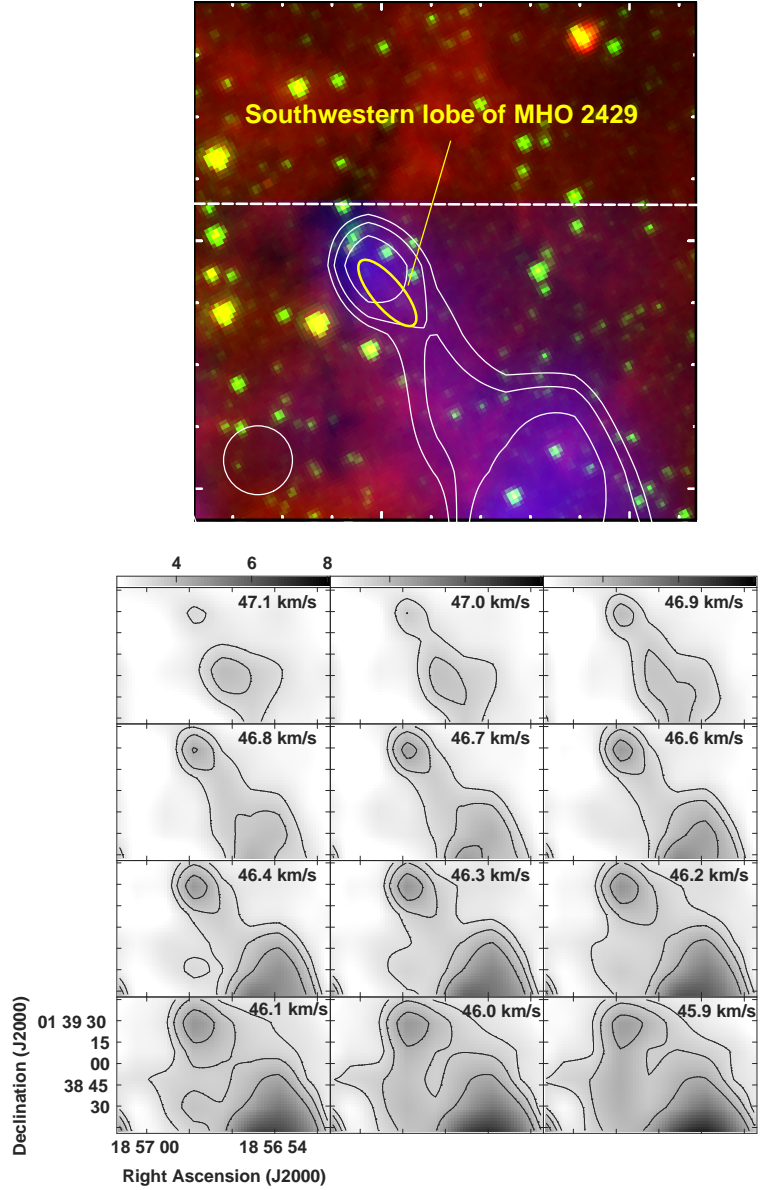}
\end{center}
\caption{Up: Same as figure \ref{outflowAll} with the $^{12}$CO J=3--2 emission
integrated between 46 and 47 km s$^{-1}$. The contour levels are 2.9, 3.1, and 3.3 K km s$^{-1}$.
The beam of the $^{12}$CO observation is included in the bottom left corner and the dashed horizontal line is the upper boundary of
the $^{12}$CO observation.
Bottom: Velocity channel maps of the $^{12}$CO J=3--2 emission showing the distribution of the molecular gas related to the H$_{2}$
southwestern lobe.
The panels show the same field as the upper image. The grayscale shown above is in K and the contour levels are 3.2, 3.6, and 4.2 K.
The rms noise of each panel is about 0.15 K.}
\label{outflow}
\end{figure}

By inspecting channel per channel the $^{12}$CO J=3--2 cube in the velocity range from 34.0 to 47.4 km s$^{-1}$,
we find that the most conspicuous CO counterpart of the H$_{2}$ emission southwest lobe peaks between 46 and 47 km s$^{-1}$.
Figure \ref{outflow} (up), shows the integration of the $^{12}$CO in this velocity range, where it can be observed an elongated clump
(see the structure delimited by the 3.1 K km s$^{-1}$~contour) aligned with the MHO southwest lobe. Towards the southwest
(bottom right corner of the image) appears another molecular clump probably not related to the outflow. Figure \ref{outflow} (bottom) displays
the emission channel maps from 45.9 to 47.1 km s$^{-1}$ in intervals of 0.11 km s$^{-1}$.
Comparing both panels of figure \ref{outflow}, it can be appreciated that the shape and alignment of the CO structure is totally in agreement with that
of the H$_{2}$ emission.
If we use the velocity range displayed in figure \ref{outflow} ($\Delta$v $\sim$ 1 km s$^{-1}$) to derive the outflow parameters,
we obtain a mass of about 9 M$_\odot$, momentum P $\sim 9~\times~{\rm cos}^{-1}(\phi)$ M$_\odot$~km s$^{-1}$,
and energy E $\sim 1 \times 10^{44}~ \times~{\rm cos}^{-2}(\phi)$ erg.
Therefore, taking into account these results and those presented above, we conclude that the parameters of the blueshifted outflow,
correlated with the southwest lobe mapped in H$_{2}$
should be within the following ranges: mass between 9 and 21 M$_\odot$, momentum: $(9-270)~\times~{\rm cos}^{-1}(\phi)$ M$_\odot$~km s$^{-1}$, and
energy:  $(1 \times 10^{44}-3.5 \times 10^{46})~\times~{\rm cos}^{-2}(\phi)$ erg.
In the following section we study the nature of the outflow driving source.

\subsection{On the Nature of the Outflow Driving Source}

\begin{figure}
\begin{center}
\FigureFile(70mm,70mm){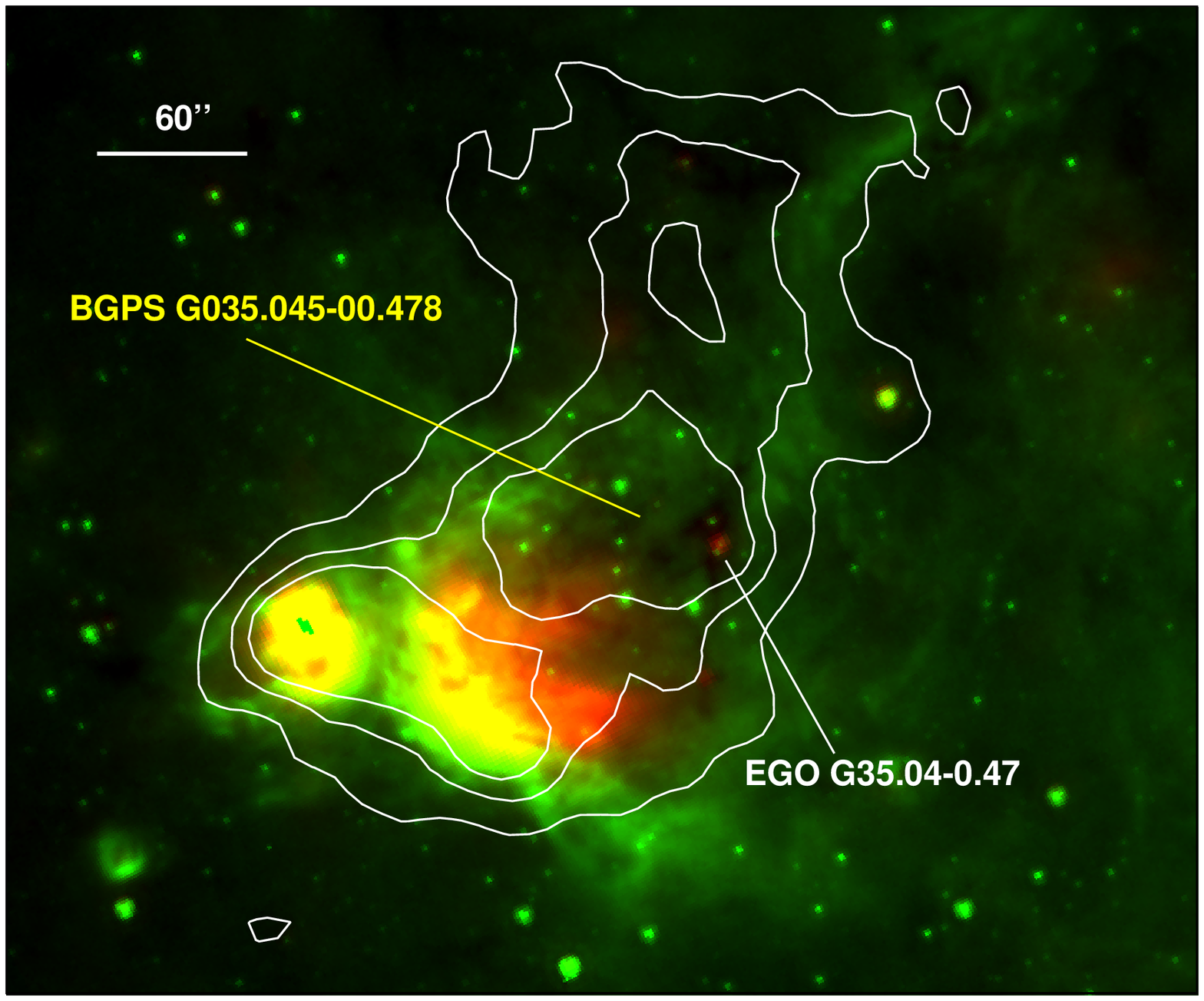}
\FigureFile(70mm,70mm){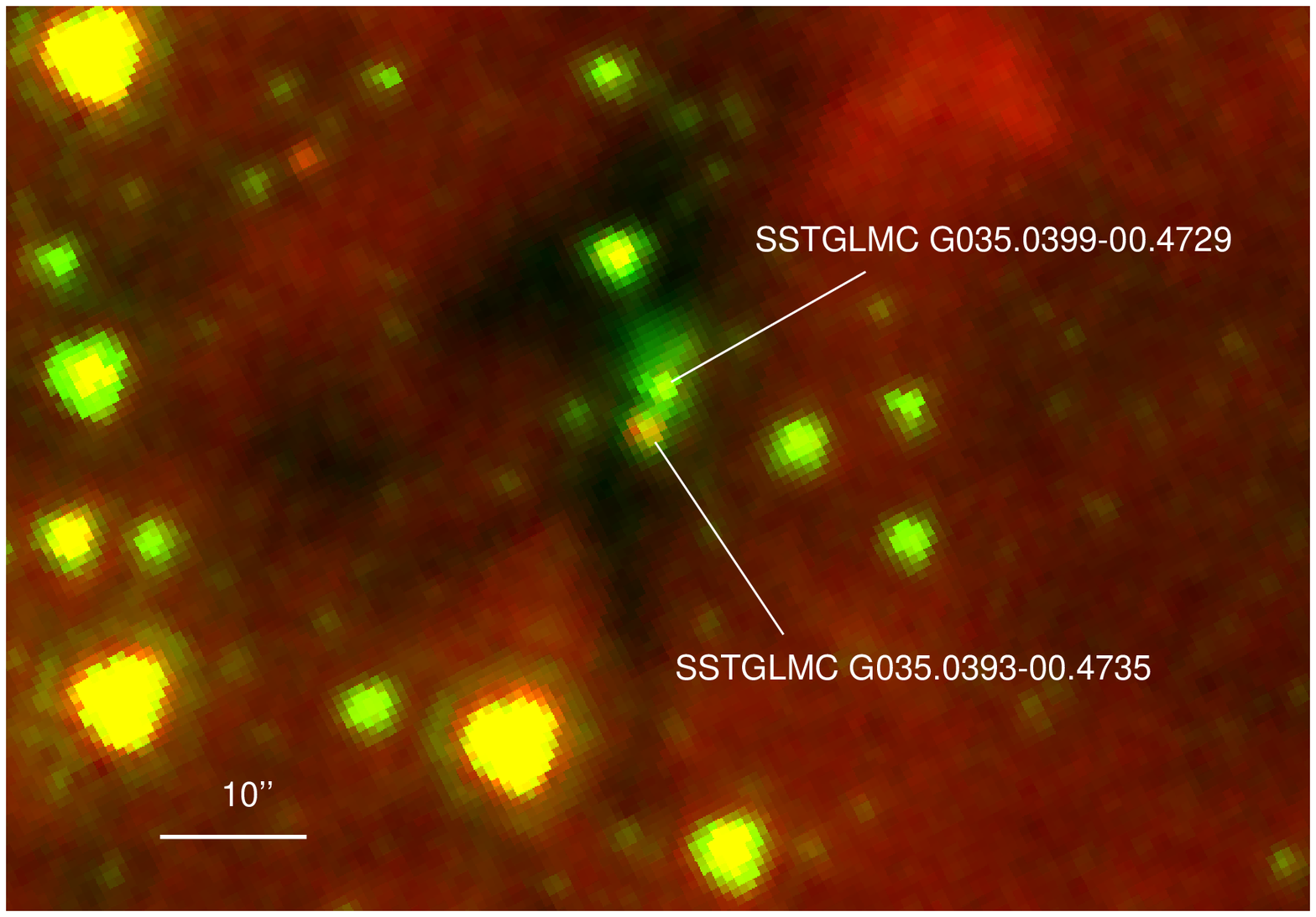}
\end{center}
\caption{Left: The 8 $\mu$m and 24 $\mu$m emissions displayed in green and red, respectively. The contours represent the smoothed
1.1 mm Bolocam emission with levels of 0.15, 0.20 and 0.30 Jy beam$^{-1}$, and the source BGPS G035.045-00.478 is indicated.
Right: 4.5 and 8 $\mu$m emissions displayed in green and red, respectively. Two Spitzer sources, from the GLIMPSE Catalog,
are marked. The source related to EGO G35.04-0.47 is the northern one.}
\label{mips}
\end{figure}

EGO G35.04-0.47 is embedded in a dusty clump mapped by the Bolocam, catalogued as source BGPS G035.045-00.478,  
which has a deconvolved radius of about 98\arcsec~\citep{roso10}.
Figure \ref{mips} (left) displays the IRAC 8 $\mu$m and the MIPS 24 $\mu$m emissions in green and red, respectively,  
where the contours are the smoothed 1.1 mm Bolocam emission. 
Using the total integrated flux S$_{1.1} = 4.71$ Jy measured towards this Bolocam source and the mass equation 
M(H$_{2}) = 14.26D^{2}S_{1.1}(e^{13/T_{d}}-1)M_{\odot}$ \citep{bally10},
where $D$ is the distance and $T_{d}$ the dust temperature assuming to be 3.4 kpc and 13 K, respectively, and applying the 
correction factor of 1.5 for the flux density \citep{aguirre11}, we obtain a total
mass of about 1900 M$_\odot$~for BGPS G035.045-00.478, indicating that the EGO is embedded in a massive cloud. 

\begin{figure}
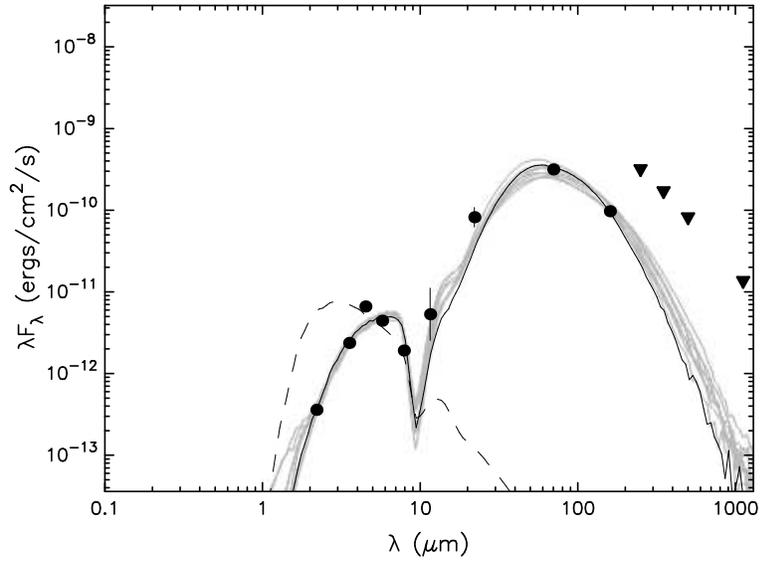

\begin{center}
\FigureFile(100mm,100mm){MHO2429_SED.eps}
\end{center}
\caption{SED of EGO G35.04-0.47. The circles
indicate the measured fluxes and the triangles are upper limits. Black and gray solid curves represent
the best-fit model and the subsequent good fittings models (with $\chi^2 - \chi^2_{best} < 3$),
respectively. The dashed line shows the stellar photosphere corresponding to the central source of the
best-fitting model, as it would look in the absence of circumstellar dust.}
\label{sed}
\end{figure}

Figure \ref{mips} (right) shows a zoom towards the EGO position displaying the IRAC 4.5 and 8.0 $\mu$m emissions in green and red, 
respectively. 
It can be seen two Spitzer sources very close each other, being the EGO counterpart
the northern one, this is the source, SSTGLMC G035.0399-00.4729.
To better characterize the nature of this source we performed a fitting of the spectral energy distribution
(SED) using the on-line tool developed by \citet{rob07}\footnote{http://caravan.astro.wisc.edu/protostars/}.
We adopt an interstellar extinction in the line of sight, $A_v$, between 4 and 50 magnitudes, and we use the distance range 3-4 kpc.
In figure \ref{sed} we show the SED with the best-fitting model (black curve), and the subsequent good fitting models (gray curves)
with $\chi^2 - \chi^2_{best} < 3$ (where $\chi^2_{best}$ is the $\chi^2$ per
data point of the best-fitting model for each source).
To construct this SED we use fluxes extracted from: UKIDSS-DR6 Galactic Plane Survey \citep{lucas08} at K band, 
source UGPS J185658.10+013936.6; GLIMPSE Source Catalog \citep{church09} at 3.6, 4.5, 5.8, and 8.0 $\mu$m, source SSTGLMC G035.0399-00.4729;
WISE All-Sky Source Catalog\footnote{WISE is a joint project of the University of California, Los Angeles, and
the Jet Propulsion Laboratory/California Institute of Technology, funded by
the NASA.} at 12 and 22~$\mu$m, 
source WISE J185658.14+013937.1; PACS (bands at 70 and 160 $\mu$m) and SPIRE (at 250, 350, and 500 $\mu$m) from Herschel, 
and finally Bolocam at 1.1 mm, source BGPS G035.045-00.478.
PACS fluxes were obtained from level 2.5 MADmaps images and SPIRE ones from level 2.5 PLW, PMW, and PSW images.
Fluxes from SPIRE and Bolocam were considered as upper limits. 

Table \ref{sedtable} 
presents a weighted mean of each constrained physical parameter
and the range arising from the good-fitting models. The weight used for the weighted means
is the inverse of the $\chi^{2}$ of each model.
The SED analysis of EGO G35.04-0.47 suggests that the central object is an intermediate mass young stellar object 
($\sim5$ M$_{\odot}$) with an envelope accretion rate of about 10$^{-4}$ M$_{\odot}$ yr$^{-1}$, similar to the mass 
accretion rates found in a sample of MYSOs by \citet{fuller05}.
Chen et al. (2010,2013) studied several EGOs and obtained mass infall rates ranging from $4 \times 10^{-2}$ 
to $1 \times 10^{-4}$ M$_{\odot}$yr$^{-1}$. 
Thus, taking into account the accretion rate, the outflow parameters found in our molecular study, and 
the presence of a massive clump of cold dust (reservoir of a large amount of matter) we suggest that the central object 
may continue to accrete mass and it probably will become a massive YSO. Finally, by considering that this YSO lays in 
the HII region G035.0-0.5, it is probable that its formation has been triggered by the expansion of the 
HII region which likely compressed a preexisting molecular clump producing its collapse.

\section{Summary}

Using new molecular line observations obtained with ASTE we studied the
ambient around the young stellar object EGO G35.04-0.47.
In particular, this EGO coincides
with the Molecular Hydrogen emission-line object MHO 2429, which is
associated with jets mapped in the H$_{2}$ near-IR emission. Taking into
account that the most conspicuous jet extends towards
the southwest, we looked for its CO counterpart. In addition, 
combining the results obtained from the ASTE data
analysis and archive images at different wavelength ranges we characterized
the outflow driving source. These results together with information of
previous studies, depict the scenario of this YSO and its possible evolution.
The main results presented in this paper are summarized below:

(a) We found that EGO G35.04-0.47 is embedded in a molecular clump mapped in the $^{12}$CO and $^{13}$CO J=3--2, and HCO$^{+}$ J=4--3 
lines peaking at about 51 km s$^{-1}$. 
This velocity is in agreement with that of a Class I 95 GHz methanol maser previously detected. 

(b) For the molecular clump we derived the abundances: X($^{12}$CO) $\sim 4.2 \times 10^{-4}$, X($^{13}$CO) $\sim 8.5 \times 10^{-6}$, 
and X(HCO$^{+}$) $\sim 1 \times 10^{-10}$.
The $^{12}$CO and $^{13}$CO abundances are in agreement with the `standard' values for dark clouds, while the
HCO$^{+}$ abundance is similar to those obtained towards some high-mass star-forming regions.

(c) Analyzing the $^{12}$CO line we discovered  high velocity molecular gas in the range from 34 to 47 km s$^{-1}$, i.e. a very likely blueshifted outflow driven 
by the EGO. The alignment and shape of this molecular structure coincide with those of the southwest lobe of MHO 2429 mainly between 
46 and 47 km s$^{-1}$, confirming that we are mapping its CO counterpart.
We estimated that the mass, momentum, and energy of this outflow should be within the following ranges: 
M $= 9-21$ M$_\odot$, P $= (9-270)~\times~{\rm cos}^{-1}(\phi)$ M$_\odot$~km s$^{-1}$, and
E$ = (1 \times 10^{44}-3.5 \times 10^{46})~\times~{\rm cos}^{-2}(\phi)$ erg, respectively, where $\phi$ is the inclination angle, which is uncertain.  

(d) Considering the results obtained from a SED analysis performed using fluxes from near-IR to mm, we suggest that the central object of EGO G35.04-0.47 
is an intermediate mass young stellar object, accreting mass at a rate similar to those found in some massive YSOs. 
Thus, taking into account the envelope accretion rate, 
the outflow parameters found in our molecular study, and the presence of a massive clump of cold dust (reservoir of a large amount of matter) 
we suggest that the central object may continue to accrete mass and it will probably become a massive YSO.

\bigskip

\noindent {\it Acknowledgements:} We would like to thank the anonymous referee
for her/his helpful suggestions and comments. 
S.P. and M.O. are members of the {\sl Carrera del 
investigador cient\'\i fico} of CONICET, Argentina. 
This work was partially supported by grants awarded by CONICET, ANPCYT and UBA (UBACyT).
M.R wishes to acknowledge support from FONDECYT (CHILE) grant No108033.
She is supported by the Chilean {\sl Center for Astrophysics} FONDAP No.
15010003. S.P. and M.O. are very grateful to the ASTE staff for the support received during the observations.
The ASTE project is driven by Nobeyama Radio Observatory (NRO), a branch of
the National Astronomical Observatory of Japan (NAOJ), in collaboration with
University of Chile, and Japanese institutes including the University of Tokyo,
Nagoya University, Osaka Prefecture University, Ibaraki University, Hokkaido
University, and Joetsu University of Education.

\eject


\newpage

\begin{table}
\caption{Parameters derived from a Gaussian fitting of the molecular spectra shown in figure \ref{spectra}.}\label{lines}
\begin{center}
\begin{tabular}{lccc}
\hline
Emission & T$_{mb}$ & V$_{\rm LSR}$ & $\Delta$v   \\
         &  (K)     & (km s$^{-1}$)      &   (km s$^{-1}$)  \\
\hline
$^{12}$CO J=3--2 & 7.0 $\pm$0.2   &  51.5 $\pm$0.1 & 4.0 $\pm$0.2 \\
                 & 2.4 $\pm$0.6   &  48.5 $\pm$0.1 & 1.0 $\pm$0.3  \\
                 & 4.4 $\pm$0.3   &  46.3 $\pm$0.2 & 3.5 $\pm$0.8  \\
                 & 2.8 $\pm$0.2   &  41.6 $\pm$0.3 & 3.8 $\pm$0.6  \\
\hline
$^{13}$CO J=3--2  & 3.6 $\pm$0.3   &  50.1 $\pm$0.1 & 2.2 $\pm$0.2   \\
                  & 1.4 $\pm$0.3  &  46.5 $\pm$0.2 & 2.0 $\pm$0.4   \\
\hline
HCO$^{+}$ J=4--3  & 1.2 $\pm$0.2 & 51.2 $\pm$0.3 & 3.0 $\pm$0.5  \\
\hline
\end{tabular}
\end{center}
\end{table}

\begin{table}
\caption{Main physical parameters from the SED of EGO G35.04-0.47.}\label{sedtable}
\begin{center}
\begin{tabular}{ccccccccc}
\hline
\multicolumn{2}{c}{$M_{\star}$}&\multicolumn{2}{c}{Age}&\multicolumn{2}{c}{$\dot
M_{env}$}&\multicolumn{2}{c}{$L$}\\
\multicolumn{2}{c}{[M$_{\odot}$]}&\multicolumn{2}{c}{[$\times
10^5$~yr]}&\multicolumn{2}{c}{ [$\times10^{-5}$ M$_{\odot}$yr$^{-1}$]}&\multicolumn{2}{c}{[$\times
10^{2} L_{\odot}$]} \\
\hline
Mean & Range & Mean & Range & Mean & Range & Mean & Range \\
5 & 3--7 & 0.8 & 0.08--4  & 9  & 3--90  & 2 & 1--4  \\
\hline
\end{tabular}
\end{center}
\end{table}

\clearpage

\clearpage

\clearpage

\clearpage

\clearpage

\clearpage

\clearpage

\clearpage

\clearpage

\clearpage

\clearpage

\begin{thebibliography}{34}
\expandafter\ifx\csname natexlab\endcsname\relax\def\natexlab#1{#1}\fi

\bibitem[{{Aguirre} {et~al.}(2011){Aguirre}, {Ginsburg}, {Dunham}, {Drosback},
  {Bally}, {Battersby}, {Bradley}, {Cyganowski}, {Dowell}, {Evans}, {Glenn},
  {Harvey}, {Rosolowsky}, {Stringfellow}, {Walawender}, \&
  {Williams}}]{aguirre11}
{Aguirre}, J.~E., {Ginsburg}, A.~G., {Dunham}, M.~K., {et~al.} 2011, \apjs,
  192, 4

\bibitem[{{Arce} {et~al.}(2007){Arce}, {Shepherd}, {Gueth}, {Lee}, {Bachiller},
  {Rosen}, \& {Beuther}}]{arce07}
{Arce}, H.~G., {Shepherd}, D., {Gueth}, F., {et~al.} 2007, Protostars and
  Planets V, 245

\bibitem[{{Bally} {et~al.}(2010){Bally}, {Aguirre}, {Battersby}, {Bradley},
  {Cyganowski}, {Dowell}, {Drosback}, {Dunham}, {Evans}, {Ginsburg}, {Glenn},
  {Harvey}, {Mills}, {Merello}, {Rosolowsky}, {Schlingman}, {Shirley},
  {Stringfellow}, {Walawender}, \& {Williams}}]{bally10}
{Bally}, J., {Aguirre}, J., {Battersby}, C., {et~al.} 2010, \apj, 721, 137

\bibitem[{{Beuther} {et~al.}(2002){Beuther}, {Schilke}, {Sridharan}, {Menten},
  {Walmsley}, \& {Wyrowski}}]{beuther02}
{Beuther}, H., {Schilke}, P., {Sridharan}, T.~K., {et~al.} 2002, \aap, 383, 892

\bibitem[{{Chen} {et~al.}(2010){Chen}, {Shen}, {Li}, {Xu}, \& {He}}]{chen10}
{Chen}, X., {Shen}, Z.-Q., {Li}, J.-J., {Xu}, Y., \& {He}, J.-H. 2010, \apj,
  710, 150

\bibitem[{{Chen} {et~al.}(2011){Chen}, {Ellingsen}, {Shen}, {Titmarsh}, \&
  {Gan}}]{chen11}
{Chen}, X., {Ellingsen}, S.~P., {Shen}, Z.-Q., {Titmarsh}, A., \& {Gan}, C.-G.
  2011, \apjs, 196, 9


\bibitem[{{Chen} {et~al.}(2013){Chen}, {Gan}, {Ellingsen}, {He}, {Shen}, \&
  {Titmarsh}}]{chen13}
{Chen}, X., {Gan}, C.-G., {Ellingsen}, S.~P., {et~al.} 2013, \apjs, 206, 9


\bibitem[{{Churchwell} {et~al.}(2009){Churchwell}, {Babler}, {Meade},
  {Whitney}, {Benjamin}, {Indebetouw}, {Cyganowski}, {Robitaille}, {Povich},
  {Watson}, \& {Bracker}}]{church09}
{Churchwell}, E., {Babler}, B.~L., {Meade}, M.~R., {et~al.} 2009, \pasp, 121,
  213

\bibitem[{{Cortes}(2011)}]{cortes11}
{Cortes}, P.~C. 2011, \apj, 743, 194

\bibitem[{{Cortes} {et~al.}(2010){Cortes}, {Parra}, {Cortes}, \&
  {Hardy}}]{cortes10}
{Cortes}, P.~C., {Parra}, R., {Cortes}, J.~R., \& {Hardy}, E. 2010, \aap, 519,
  A35

\bibitem[{{Cutri} {et~al.}(2012){Cutri}, {Wright}, {Conrow}, {Bauer},
  {Benford}, {Brandenburg}, {Dailey}, {Eisenhardt}, {Evans}, {Fajardo-Acosta},
  {Fowler}, {Gelino}, {Grillmair}, {Harbut}, {Hoffman}, {Jarrett},
  {Kirkpatrick}, {Leisawitz}, {Liu}, {Mainzer}, {Marsh}, {Masci}, {McCallon},
  {Padgett}, {Ressler}, {Royer}, {Skrutskie}, {Stanford}, {Wyatt}, {Tholen},
  {Tsai}, {Wachter}, {Wheelock}, {Yan}, {Alles}, {Beck}, {Grav}, {Masiero},
  {McCollum}, {McGehee}, {Papin}, \& {Wittman}}]{cutri12}
{Cutri}, R.~M., {Wright}, E.~L., {Conrow}, T., {et~al.} 2012, {Explanatory
  Supplement to the WISE All-Sky Data Release Products}, Tech. rep.

\bibitem[{{Cyganowski} {et~al.}(2013){Cyganowski}, {Koda}, {Rosolowsky},
  {Towers}, {Donovan Meyer}, {Egusa}, {Momose}, \& {Robitaille}}]{cyga2013}
{Cyganowski}, C.~J., {Koda}, J., {Rosolowsky}, E., {et~al.} 2013, \apj, 764, 61

\bibitem[{{Cyganowski} {et~al.}(2011){Cyganowski}, {Brogan}, {Hunter},
  {Churchwell}, \& {Zhang}}]{cyga11}
{Cyganowski}, C.~J., {Brogan}, C.~L., {Hunter}, T.~R., {Churchwell}, E., \&
  {Zhang}, Q. 2011, \apj, 729, 124


\bibitem[{{Cyganowski} {et~al.}(2008){Cyganowski}, {Whitney}, {Holden},
  {Braden}, {Brogan}, {Churchwell}, {Indebetouw}, {Watson}, {Babler},
  {Benjamin}, {Gomez}, {Meade}, {Povich}, {Robitaille}, \& {Watson}}]{cyga08}
{Cyganowski}, C.~J., {Whitney}, B.~A., {Holden}, E., {et~al.} 2008, \aj, 136,
  2391

\bibitem[{{Davis} {et~al.}(2010){Davis}, {Gell}, {Khanzadyan}, {Smith}, \&
  {Jenness}}]{davis10}
{Davis}, C.~J., {Gell}, R., {Khanzadyan}, T., {Smith}, M.~D., \& {Jenness}, T.
  2010, \aap, 511, A24

\bibitem[{{De Buizer} \& {Vacca}(2010)}]{debuizer10}
{De Buizer}, J.~M. \& {Vacca}, W.~D. 2010, \aj, 140, 196

\bibitem[{{Dirienzo} {et~al.}(2012){Dirienzo}, {Indebetouw}, {Brogan},
  {Cyganowski}, {Churchwell}, \& {Friesen}}]{diri12}
{Dirienzo}, W.~J., {Indebetouw}, R., {Brogan}, C., {et~al.} 2012, \aj, 144, 173


\bibitem[{{Ezawa} {et~al.}(2004){Ezawa}, {Kawabe}, {Kohno}, \&
  {Yamamoto}}]{ezawa04}
{Ezawa}, H., {Kawabe}, R., {Kohno}, K., \& {Yamamoto}, S. 2004, in Presented at
  the Society of Photo-Optical Instrumentation Engineers (SPIE) Conference,
  Vol. 5489, Society of Photo-Optical Instrumentation Engineers (SPIE)
  Conference Series, ed. J.~M. {Oschmann}, Jr., 763--772

\bibitem[{{Falgarone} {et~al.}(1994){Falgarone}, {Lis}, {Phillips}, {Pouquet},
  {Porter}, \& {Woodward}}]{falgarone94}
{Falgarone}, E., {Lis}, D.~C., {Phillips}, T.~G., {et~al.} 1994, \apj, 436, 728

\bibitem[{{Froebrich} {et~al.}(2003){Froebrich}, {Smith}, {Hodapp}, \&
  {Eisl{\"o}ffel}}]{froebrich03}
{Froebrich}, D., {Smith}, M.~D., {Hodapp}, K.-W., \& {Eisl{\"o}ffel}, J. 2003,
  \mnras, 346, 163

\bibitem[{{Fuller} {et~al.}(2005){Fuller}, {Williams}, \&
  {Sridharan}}]{fuller05}
{Fuller}, G.~A., {Williams}, S.~J., \& {Sridharan}, T.~K. 2005, \aap, 442, 949


\bibitem[{{He} {et~al.}(2012){He}, {Takahashi}, \& {Chen}}]{he12}
{He}, J.~H., {Takahashi}, S., \& {Chen}, X. 2012, \apjs, 202, 1

\bibitem[{{Kuchar} \& {Clark}(1997)}]{kuchar97}
{Kuchar}, T.~A. \& {Clark}, F.~O. 1997, \apj, 488, 224

\bibitem[{{Lawrence} {et~al.}(2007){Lawrence}, {Warren}, {Almaini}, {Edge},
  {Hambly}, {Jameson}, {Lucas}, {Casali}, {Adamson}, {Dye}, {Emerson},
  {Foucaud}, {Hewett}, {Hirst}, {Hodgkin}, {Irwin}, {Lodieu}, {McMahon},
  {Simpson}, {Smail}, {Mortlock}, \& {Folger}}]{law07}
{Lawrence}, A., {Warren}, S.~J., {Almaini}, O., {et~al.} 2007, \mnras, 379,
  1599


\bibitem[{{Lee} {et~al.}(2012){Lee}, {Takami}, {Duan}, {Karr}, {Su}, {Liu},
  {Froebrich}, \& {Yeh}}]{lee12}
{Lee}, H.-T., {Takami}, M., {Duan}, H.-Y., {et~al.} 2012, \apjs, 200, 2

\bibitem[{{Lockman}(1989)}]{lockman89}
{Lockman}, F.~J. 1989, \apjs, 71, 469

\bibitem[{{Lucas} {et~al.}(2008){Lucas}, {Hoare}, {Longmore}, {Schr{\"o}der},
  {Davis}, {Adamson}, {Bandyopadhyay}, {de Grijs}, {Smith}, {Gosling},
  {Mitchison}, {G{\'a}sp{\'a}r}, {Coe}, {Tamura}, {Parker}, {Irwin}, {Hambly},
  {Bryant}, {Collins}, {Cross}, {Evans}, {Gonzalez-Solares}, {Hodgkin},
  {Lewis}, {Read}, {Riello}, {Sutorius}, {Lawrence}, {Drew}, {Dye}, \&
  {Thompson}}]{lucas08}
{Lucas}, P.~W., {Hoare}, M.~G., {Longmore}, A., {et~al.} 2008, \mnras, 391, 136

\bibitem[{{Milam} {et~al.}(2005){Milam}, {Savage}, {Brewster}, {Ziurys}, \&
  {Wyckoff}}]{milam05}
{Milam}, S.~N., {Savage}, C., {Brewster}, M.~A., {Ziurys}, L.~M., \& {Wyckoff},
  S. 2005, \apj, 634, 1126

\bibitem[{{Mottram} \& {Brunt}(2012)}]{mottram12}
{Mottram}, J.~C. \& {Brunt}, C.~M. 2012, \mnras, 420, 10

\bibitem[{{Noriega-Crespo} {et~al.}(2004){Noriega-Crespo}, {Morris}, {Marleau},
  {Carey}, {Boogert}, {van Dishoeck}, {Evans}, {Keene}, {Muzerolle},
  {Stapelfeldt}, {Pontoppidan}, {Lowrance}, {Allen}, \& {Bourke}}]{noriega04}
{Noriega-Crespo}, A., {Morris}, P., {Marleau}, F.~R., {et~al.} 2004, \apjs,
  154, 352


\bibitem[{{Ortega} {et~al.}(2012){Ortega}, {Paron}, {Cichowolski}, {Rubio}, \&
  {Dubner}}]{ortega12}
{Ortega}, M.~E., {Paron}, S., {Cichowolski}, S., {Rubio}, M., \& {Dubner}, G.
  2012, \aap, 546, A96

\bibitem[{{Paron} {et~al.}(2011){Paron}, {Petriella}, \& {Ortega}}]{paron11}
{Paron}, S., {Petriella}, A., \& {Ortega}, M.~E. 2011, \aap, 525, A132

\bibitem[{{Peretto} \& {Fuller}(2009)}]{peretto09}
{Peretto}, N. \& {Fuller}, G.~A. 2009, \aap, 505, 405

\bibitem[{{Robitaille} {et~al.}(2007){Robitaille}, {Whitney}, {Indebetouw}, \&
  {Wood}}]{rob07}
{Robitaille}, T.~P., {Whitney}, B.~A., {Indebetouw}, R., \& {Wood}, K. 2007,
  \apjs, 169, 328

\bibitem[{{Rosolowsky} {et~al.}(2010){Rosolowsky}, {Dunham}, {Ginsburg},
  {Bradley}, {Aguirre}, {Bally}, {Battersby}, {Cyganowski}, {Dowell},
  {Drosback}, {Evans}, {Glenn}, {Harvey}, {Stringfellow}, {Walawender}, \&
  {Williams}}]{roso10}
{Rosolowsky}, E., {Dunham}, M.~K., {Ginsburg}, A., {et~al.} 2010, \apjs, 188,
  123

\bibitem[{{Schlingman} {et~al.}(2011){Schlingman}, {Shirley}, {Schenk},
  {Rosolowsky}, {Bally}, {Battersby}, {Dunham}, {Ellsworth-Bowers}, {Evans},
  {Ginsburg}, \& {Stringfellow}}]{schlin11}
{Schlingman}, W.~M., {Shirley}, Y.~L., {Schenk}, D.~E., {et~al.} 2011, \apjs,
  195, 14

\bibitem[{{Scoville} {et~al.}(1986){Scoville}, {Sargent}, {Sanders},
  {Claussen}, {Masson}, {Lo}, \& {Phillips}}]{scoville86}
{Scoville}, N.~Z., {Sargent}, A.~I., {Sanders}, D.~B., {et~al.} 1986, \apj,
  303, 416

\bibitem[{{Ukidss}(2012)}]{ukidss12}
{Ukidss}, C. 2012, VizieR Online Data Catalog, 2316, 0

\bibitem[{{Urquhart} {et~al.}(2009){Urquhart}, {Hoare}, {Purcell}, {Lumsden},
  {Oudmaijer}, {Moore}, {Busfield}, {Mottram}, \& {Davies}}]{urqu09}
{Urquhart}, J.~S., {Hoare}, M.~G., {Purcell}, C.~R., {et~al.} 2009, \aap, 501,
  539

\bibitem[{{van Dishoeck} \& {Hogerheijde}(1999)}]{vandish99}
{van Dishoeck}, E.~F. \& {Hogerheijde}, M.~R. 1999, in NATO ASIC Proc. 540: The
  Origin of Stars and Planetary Systems, ed. C.~J. {Lada} \& N.~D. {Kylafis},
  97

\bibitem[{{Wu} {et~al.}(2004){Wu}, {Wei}, {Zhao}, {Shi}, {Yu}, {Qin}, \&
  {Huang}}]{wu04}
{Wu}, Y., {Wei}, Y., {Zhao}, M., {et~al.} 2004, \aap, 426, 503

\end{thebibliography}
\end{document}